\documentclass[preprint]{aastex}

\usepackage{natbib,graphicx,textcomp}

\shorttitle{transit timing monitoring of HAT-P-13b}
\shortauthors{Fulton et al.}

\begin{document}

\title{Long-Term Transit Timing Monitoring and Refined Light Curve Parameters of HAT-P-13b}

\author{Benjamin J.~Fulton\altaffilmark{1},
Avi Shporer\altaffilmark{1, 2},
Joshua N.~Winn\altaffilmark{3},
Matthew J.~Holman\altaffilmark{4},
Andr\'as P\'al\altaffilmark{5,6},
J. Zachary Gazak\altaffilmark{7}
}


\altaffiltext{1}{Las Cumbres Observatory Global Telescope Network, 6740 Cortona Drive, Suite 102, Santa Barbara, CA 93117, USA; bjfulton@lcogt.net}
\altaffiltext{2}{Department of Physics, Broida Hall, University of California, Santa Barbara, CA 93106, USA}
\altaffiltext{3}{Department of Physics and Kavli Institute for Astrophysics and Space Research, Massachusetts Institute of Technology, Cambridge, MA 02139, USA}
\altaffiltext{4}{Harvard-Smithsonian Center for Astrophysics, Cambridge, MA 02138, USA}
\altaffiltext{5}{Konkoly Observatory of the Hungarian Academy of Sciences, Konkoly Thege Mikl\'os \'ut 15-17, Budapest H-1121, Hungary}
\altaffiltext{6}{Department of Astronomy, Lor\'and E\"otv\"os University, P\'azm\'any P\'eter s\'et\'any 1/A, Budapest H-1117, Hungary}
\altaffiltext{7}{Institute for Astronomy, University of Hawaii, 2680 Woodlawn Dr, Honolulu, HI 96822}

\begin{abstract}

We present 10 new transit light curves of the transiting hot Jupiter HAT-P-13b, obtained during two observational seasons by three different telescopes. When combined with 12 previously published light curves, we have a sample consisting of 22 transit light curves, spanning 1,041 days across four observational seasons. We use this sample to examine the recently observed large-amplitude transit timing variations \citep{Pal11}, and give refined system parameters. We find that the transit times are consistent with a linear ephemeris, with the exception of a single transit time, from UT 2009 Nov 5, for which the measured mid transit time significantly deviates from our linear ephemeris. The nature of this deviation is not clear, and the rest of the data do not show any significant transit timing variation.

\end{abstract}

\section{Introduction}

HAT-P-13 \citep{Bakos09} is among the brightest stars ($V$=10.6 mag) hosting a multi-planet system containing a transiting planet, HAT-P-13b ($M_{p,b} = 0.85\ M_J$, $R_{p,b} = 1.3\ R_J$, \citealt{Bakos09, Winn10}). A transit depth of $\sim$1\% and a short orbital period of $\sim$2.92 days allow for many transits that can be observed by small ground-based telescopes, making this system a good subject of observational studies related to planetary systems \citep[e.g.,][]{Mardling10, Winn10, Payne11}.

HAT-P-13b was discovered as a transiting planet by \cite{Bakos09}, who also identified a second planet in the system, HAT-P-13c ($M_{p,c}\sin i_c = 14.3\ M_J$, \citealt{Winn10}), moving in an eccentric orbit ($e\approx0.7$), with a period of about 1.2 years. \cite{Winn10} gathered additional radial velocity (RV) measurements and identified a third low mass companion in the system, possibly a third planet, whose period is currently unknown but expected to be a few years or longer. In addition, \cite{Winn10} identified that HAT-P-13b orbit is likely to be aligned with the host star's equator. As indicated by \cite{Winn10}, based on the analysis of \cite{Mardling10}, a spin-orbit alignment of HAT-P-13b suggests a small mutual orbital inclination of planets b and c, suggesting in turn the possibility that planet c is also transiting. So far no transits of HAT-P-13c have been detected, although some attempts were made to look for it during the 2010 predicted conjunction time \citep{Szabo10}.

In a recent paper \cite{Pal11} analyzed transit timing of HAT-P-13b from four observational seasons, 2007/2008 (hereafter Season 1), 2008/2009 (hereafter Season 2), 2009/2010 (hereafter Season 3) and 2010/2011 (hereafter Season 4), and identified a deviation of the transit times from the predicted times during the last season, of about 0.015 day. In principle, this kind of long term transit timing variation (TTV) could be due to the presence of another planet in the system, in a large eccentric orbit, as described for example by \citet[][their Section 4]{Agol05}. The known orbit of HAT-P-13c does not match the TTV pattern identified by \cite{Pal11}, but it could be due to the third planet suggested by \cite{Winn10}, or a further companion in the system.

We have set out here to study the suggested TTV signal in more detail. We present 10 new HAT-P-13b transit light curves, 5 from each of the Seasons 3 and 4, and combine them with the 12 light curves which were available to \cite{Pal11}. Therefore, our analysis is based on a total of 22 transit light curves, either partial or complete, from four consecutive observational seasons, including a single light curve form Season 1 and 6--8 light curves per season for Seasons 2--4. Overall, our data span 1,041 days. Obtaining the new data and photometric processing is described in Section 2, and in Section 3 we present our transit light curve analysis. We discuss our results in Section 4, and give a short summary in Section 5.

\section{Observations}
\label{obs}

Our 10 new transit light curves of HAT-P-13b from Seasons 3 and 4 were obtained at three observatories. A brief description of the three telescopes and instruments used is given in the following paragraphs.

\emph{Fred Lawrence Whipple Observatory (FLWO):} FLWO is located on Mount Hopkins, near Amado, AZ.
We used the KeplerCam 4k$\times$4k Fairchild CCD486 mounted
on the FLWO 1.2m telescope. KeplerCam has a pixel scale of $0\farcs{62}$ pixel$^{-1}$ (2$\times$2 binning), and a $23\farcm{1}\times23\farcm{1}$  field of view (FOV). All four new FLWO light curves, two from each of Seasons 3 and 4, were obtained in the SDSS-$i'$ filter. These observations were conducted with the telescope nominally in focus, but the optical characteristics of the telescope create a relatively large point spread function (PSF).
The seven light curves from the discovery paper \citep{Bakos09} were also obtained with the FLWO 1.2m and KeplerCam, and here we used a similar setup. 

\emph{Faulkes Telescope North (FTN):} FTN is located on Mauna Haleakala in Maui, HI.
We obtained four light curves from LCOGT's robotic 2.0 m telescope using the Spectral Instruments camera
and a Pan-STARRS Z filter. The camera consists of a 4k$\times$4k Fairchild Imaging CCD with a pixel scale
of $0\farcs{304}$ pixel$^{-1}$ (2$\times$2 binning) and a FOV of $10\farcm{5}\times10\farcm{5}$. Exposure times ranged from 6 s to 10 s and a slight defocus was applied to the telescope in order to project the PSF onto a larger number of
pixels, prevent saturation and increase the open shutter time relative to the overall cycle time.
Three light curves were obtained during Season 3, and one during Season 4.

\emph{Byrne Observatory at Sedgwick (BOS):} BOS\footnote{Located at: 34.687604\textdegree, -120.039067\textdegree, 500m} is located at the Sedgwick Reserve near Santa Ynez, CA. We obtained two transit light curves of HAT-P-13b, both during Season 4, using the RC Optics 0.8 m remotely operated telescope at BOS. This telescope is equipped with
a Santa Barbara Instrument Group (SBIG) STL-6303E camera containing a 3k$\times$2k Kodak Enhanced KAF-6306E CCD with a
pixel scale of $0\farcs{572}$ pixel$^{-1}$ (2$\times$2 binning) and a $14\farcm{7}\times9\farcm{8}$ FOV. We observed in the SDSS-$i'$ filter, and exposure times ranged from 50 s to 80 s depending on atmospheric conditions. Due to the smaller aperture and short readout time ($\sim$10 s) at BOS, no defocusing was applied.

\begin{figure}[h]
\epsscale{0.93}
\plotone{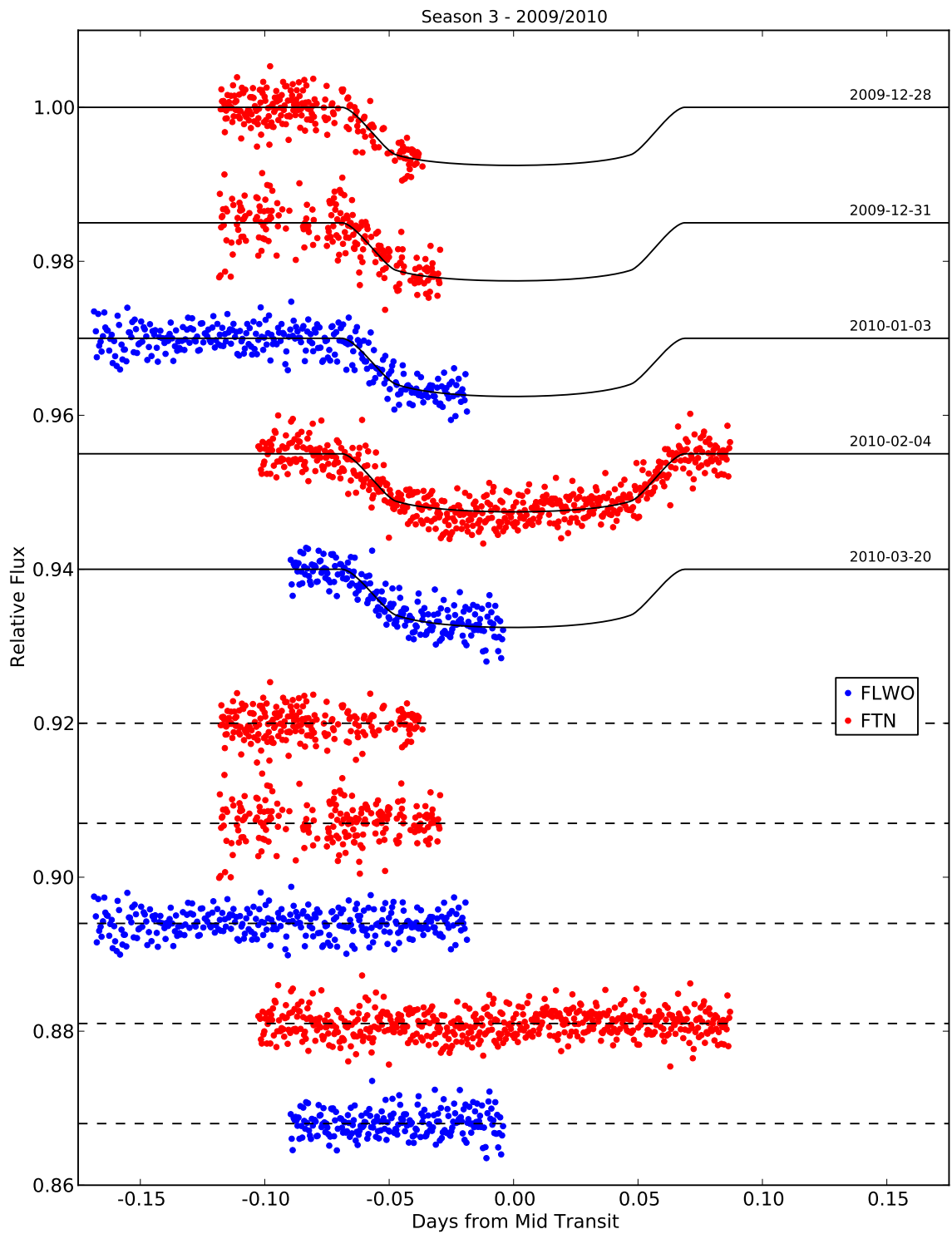}
\centering
\caption{The five new light curves obtained in this work during Season 3, plotted in chronological order starting from the top, and offset in relative flux for clarity. The residuals after model subtraction appear below in the same order. The best fitting model, from the analysis of all light curves, is over-plotted in black. All dates are UT at the start of observation.}
\label{lc1}
\end{figure}

\begin{figure}[h]
\epsscale{0.93}
\plotone{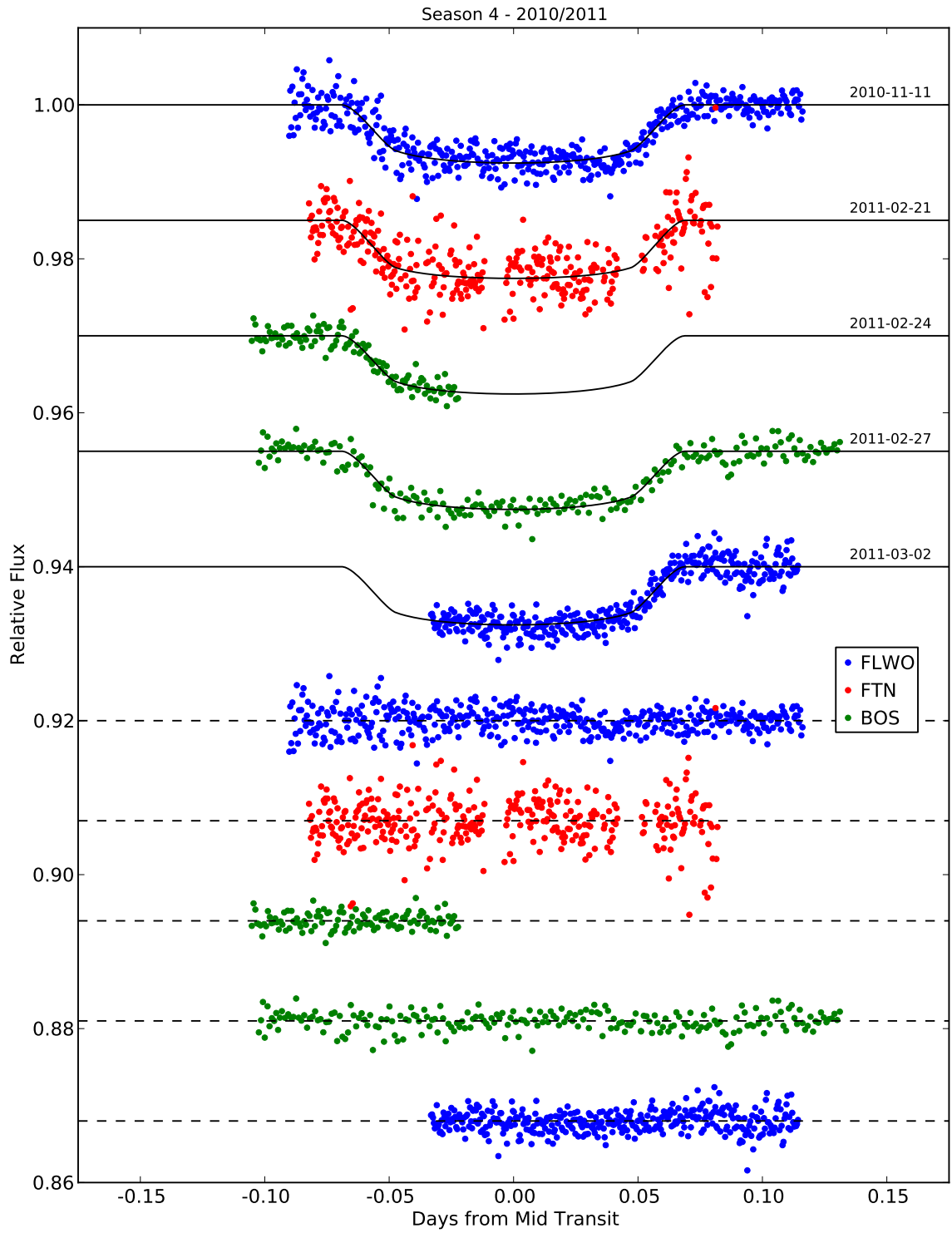}
\centering
\caption{Same as Figure \ref{lc1}, but for Season 4.}
\label{lc2}
\end{figure}

At all observatories we gathered CCD images encompassing the target star HAT-P-13. The moderately populated 
field surrounding the target provided several stars of similar brightness within the FOVs, to be used as comparison stars in the photometric processing. All data were reduced using standard routines for bias subtraction, dark current subtraction
(when necessary), and flat-field correction. We extracted light curves with PyRAF using aperture photometry by dividing the flux of the target
star by the weighted summed flux of several comparison stars in each image. Julian Dates of mid exposure were recorded during the observations, and later converted to BJD\_TDB using the tools described in \cite{Eastman10}\footnote{Online tool for HJD\_UTC to BJD\_TDB conversion; http://astroutils.astronomy.ohio-state.edu/time/hjd2bjd.html}. We optimized aperture sizes and the selection of comparison stars by minimizing the scatter of the resulting light curves, while iteratively removing 5 $\sigma$ outliers. A total of 2 photometric outlier data points were removed from the collection of all light curves. All 10 new light curves are shown in Figures~\ref{lc1} and \ref{lc2}, and are listed in Table~\ref{lclist}.

In addition to the 10 new light curves we obtained for this work, we also re-analyzed 12 light curves available in the literature. Those include one light curve from Season 1 and six from Season 2, all from \cite{Bakos09}, two light curves presented by \cite{Szabo10} from Season 3, and three light curves from \cite{Pal11} from Season 4. We adopted the previously published light curves as they were presented in their respective papers, and we only redid the fits in this work. Table~\ref{obstable} lists all 22 transit light curves included in our analysis.

\section{Analysis}
\label{analysis}

\subsection{Analysis of all available data}

Our light curve fitting was done using the Transit Analysis Package\footnote{http://ifa.hawaii.edu/users/zgazak/IfA/TAP.html} \citep[TAP;][]{Gazak11}. TAP utilizes Monte Carlo Markov Chains (MCMC) with the Metropolis-Hastings algorithm and a Gibbs sampler \citep[e.g.,][]{Ford05, Ford06, Holman06, Cameron07, Burke07}. To account for possible temporally correlated noise \citep[e.g.,][]{Pont06} TAP uses the wavelet likelihood approach of \cite{Carter09}. TAP has the ability to simultaneously fit 13 parameters: orbital period ($P$), mid transit time ($T_c$), orbital inclination ($i$), orbital semi-major axis normalized by the host star's radius ($a/R_{s}$), planet to star radii ratio ($R_p/R_{s}$), two limb darkening coefficients ($u_1$ and $u_2$) for a quadratic limb darkening law, orbital eccentricity ($e$) and longitude of periastron ($\omega$). In addition, TAP fits a linear slope ($S$), to account for a possible linear trend with time during the transit, a flux normalization factor ($N$), and two noise components: a temporally uncorrelated Gaussian ``white" noise ($\sigma_{w}$) and a time correlated ``red" noise ($\sigma_{r}$) \cite[see equations 32--34 of][]{Carter09}, where a power spectrum density of $1/f$ is assumed.

We determined limb-darkening coefficients\footnote{$u_{1,V}$=0.5162 $u_{2,V}$=0.2448, $u_{1,R}$=0.3971 $u_{2,R}$=0.2977, $u_{1,I}$=0.2922 $u_{2,I}$=0.3192, $u_{1,i'}$=0.3208 $u_{2,i'}$=0.3124, $u_{1,Z}$=0.2441 $u_{2,Z}$=0.3226}
by interpolating over the grids of \cite{Claret00,Claret04} and fixed these parameters in the analysis. Since $e$ and $\omega$ are not well constrained by light curves alone Gaussian priors were assigned to these parameters using the values from \citet{Winn10}: $e=0.0133\pm0.0041$, and $\omega=210^{+27}_{-36}$ degrees. Therefore our model includes five parameters simultaneously fitted to all light curves ($P$, $T_c$, $i$, $a/R_s$, and $R_p/R_s$), and 22 sets of four parameters ($S$, $N$, $\sigma_{w}$, and $\sigma_{r}$) fitted individually to each light curve.
We used jump rates of 25\% for all free MCMC parameters, and ran 10 chains of 10$^{5}$ steps each, discarding the first 10\% of each chain before combining results of all chains. Each chain started from a different initial position 10 $\sigma$ away from the optimized parameter values. The best fit values, and upper and lower 1 $\sigma$ errors for each parameter were determined by taking the median, 15.9, and 85.1 percentile values respectively of the resulting a posteriori probability distributions. In order to check the chains for non-convergence, we calculated the Gelman-Rubin statistic \citep{Gelman03,Ford06,Holman06}. The ratio
of interchain variance to the intrachain variance was found to be within 10\% of unity for each free parameter, giving no indication of non-convergence.
Results of this analysis are shown in the bold row of Table \ref{params}, and we used those parameters for the over-plotted model in Figures~\ref{lc1}--\ref{szabo}.

Table \ref{obstable} includes parameters that indicate the quality of each light curve. The photometric noise rate (PNR) is defined as PNR=RMS/$\sqrt{\Gamma}$, where the root mean square (RMS) is derived from the light curve residuals and $\Gamma$ is the median number of cycles (including exposure time and any dead time such as readout time) per minute. Also listed are $\sigma_{w}$ and $\sigma_{r}$, as fitted by the TAP.

\subsection{Seasonal Analyses}

We repeated this process treating the collection of all light curves from each of the three seasons 2--4 as separate subsets in order to look for possible variations in the system parameters from season to season. The resulting system parameters determined from all light curves and the three seasonal analyses can be found in Table \ref{params}. The bottom line of the table lists the parameters from \cite{Bakos09}, for comparison.

\subsection{Refined Ephemeris}

We used the results from fitting all light curves in order to look for TTV and determine a refined ephemeris. For that end we analyzed each light curve separately by allowing only the mid transit time and the four light curve specific parameters ($S$, $N$, $\sigma_{w}$, and $\sigma_{r}$) to vary. The resulting mid-transit times for each transit event are listed in Table \ref{midtable}. 

Once we determined the mid-transit times and the errors on those measurements,
we then performed a linear least squares fit for a linear ephemeris, including $P$ and a reference epoch $T_{c,0}$. Since we have some freedom in choosing the epoch for which $T_{c,0}$ is fitted, we chose it to be during season 3, when the covariance between $P$ and $T_{c,0}$ is minimized, although we do not have a light curve of that specific transit event. The resulting parameters and their uncertainties are listed in the bold row of Table \ref{params}. We verified that the resulting cov($P$, $T_{c,0}$) is small enough and can be neglected when propagating the error bars to future (or past) mid transit times.

\section{Discussion}
\label{discussion}

The transit times O$-$C diagram showing the residuals from our linear ephemeris is presented in Figure~\ref{oc}, and the residuals are listed in Table~\ref{midtable} as time difference, in seconds, and also after dividing by the mid transit times uncertainty, to show the significance of the difference. A close look at Figure~\ref{oc} shows there is only a single significant outlier (12.7 min, 5.2 $\sigma$), the earlier of the two transit events obtained by \cite{Szabo10} during Season 3, on UT 2009 Nov 5.
The linear fit to the mid-transit times produced $\chi^{2}=45.57$ with 20 degrees of freedom (DoF), and a reduced $\chi^2$ of
$\chi_{red}^{2}=2.28$. However, this value
is highly affected by the $\sim$5 $\sigma$ outlier from the UT 2009 Nov 5 transit. If this single point is ignored we get $\chi^{2}=19.36$ with 19 DoF, and $\chi_{red}^{2}=1.02$, but the difference in the resulting fit is small ($T_{c,0}$ and $P$ changed by $-0.5$ $\sigma$ and 0.1 $\sigma$, respectively). 
The RMS of the O$-$C residuals
including the UT 2009 Nov 5 event is 211 s, and 144 s without including that event. 

Comparing our O$-$C diagram to the one presented by \citet[][see their Figure 2]{Pal11} shows a dramatic difference. Their figure shows that the mid transit times of the three light curves they obtained, during Season 4, strongly deviate from a linear ephemeris, by about 0.015 days, or 3--18 $\sigma$ according to the mid transit time uncertainties they provide. The linear ephemeris derived in \citet{Szabo10} and adopted by \cite{Pal11} was based only on the seven \cite{Bakos09} transit times, from Seasons 1 and 2, and the two from \cite{Szabo10}, from Season 3. Therefore the UT 2009 Nov 5 transit time from Season 3 heavily affected their derived ephemeris.

 Here we have seven events from Season 3, including three observed by the FTN 2.0 m and two by the FLWO 1.2 m. Our larger number of observed transits from that season suggests that the UT 2009 Nov 5 mid transit time measurement is a single outlier, and that the data we have at hand are consistent with a linear ephemeris.

\begin{figure}[h]
\epsscale{1.1}
\plottwo{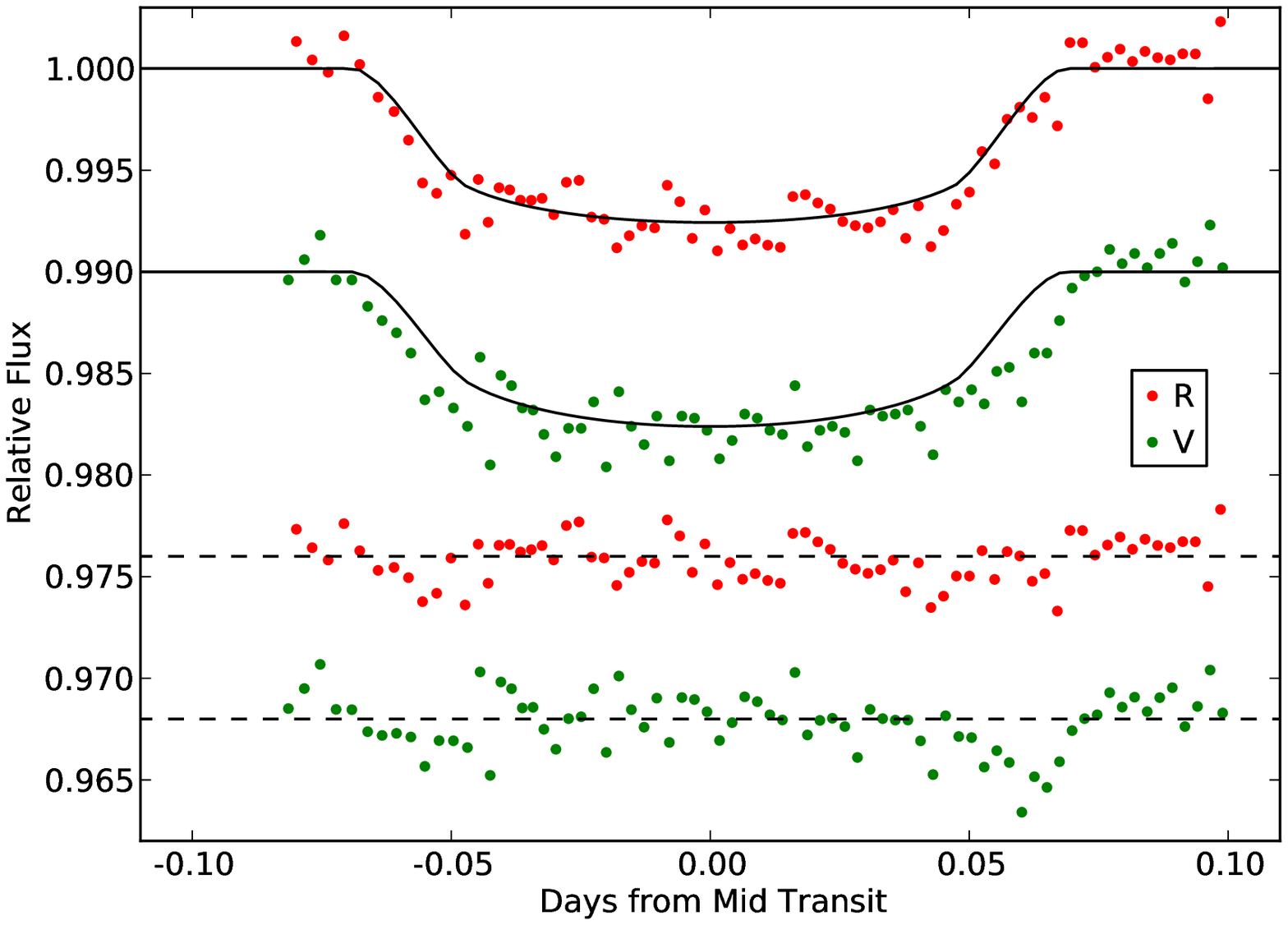}{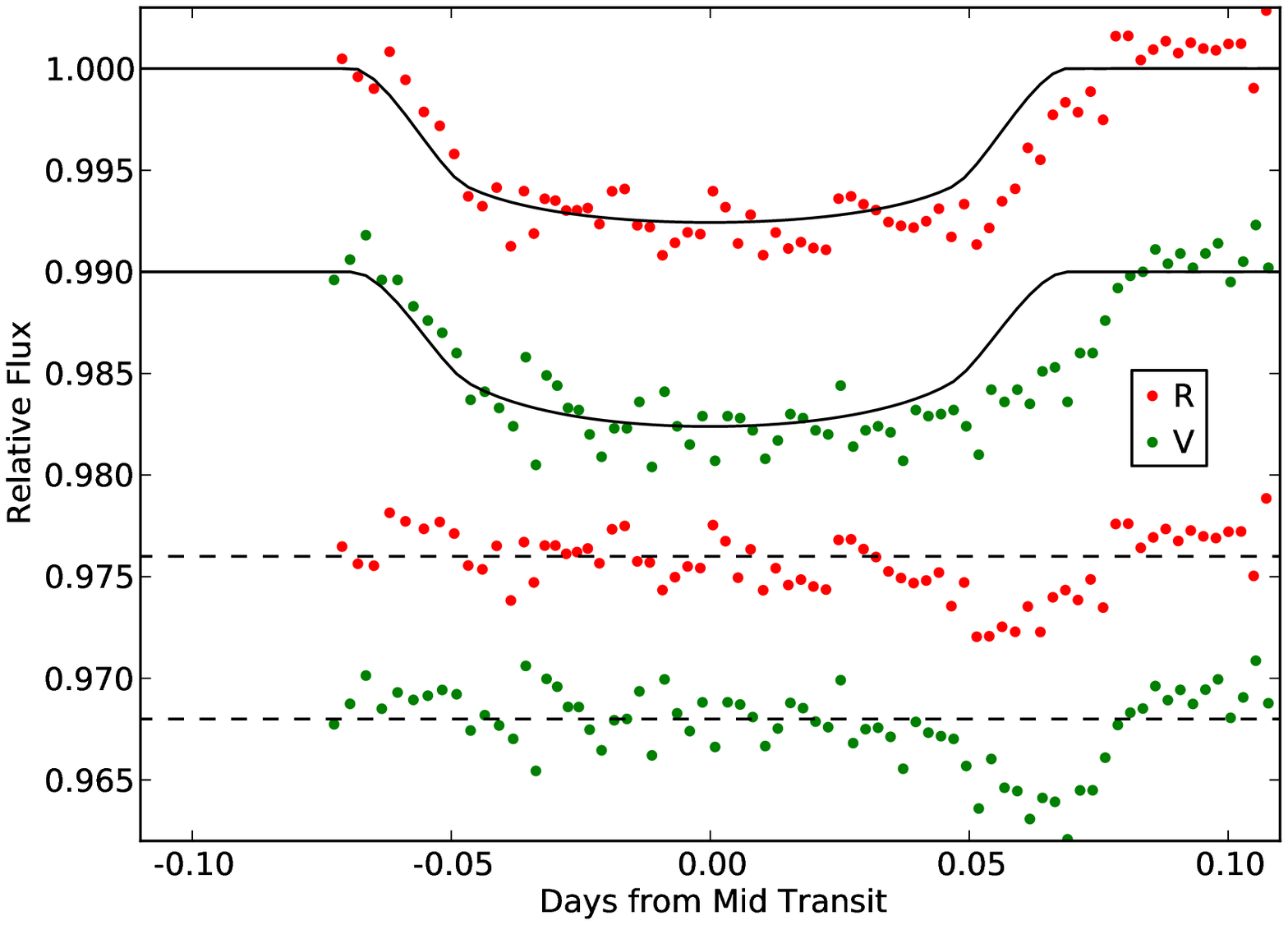}
\centering
\caption{The UT 2009 Nov 5 light curve, which was obtained by \citet{Szabo10}. Our best-fitting model, obtained from TAP analysis of all 22 light curves is over-plotted in black. \emph{Left:} The model is phased and shifted to the best-fitting
mid-transit time for this event. \emph{Right:} The model is phased and shifted to the best-fitting linear ephemeris from Table \ref{params}.}
\label{szabo}
\end{figure}

\begin{figure}[h]
\includegraphics{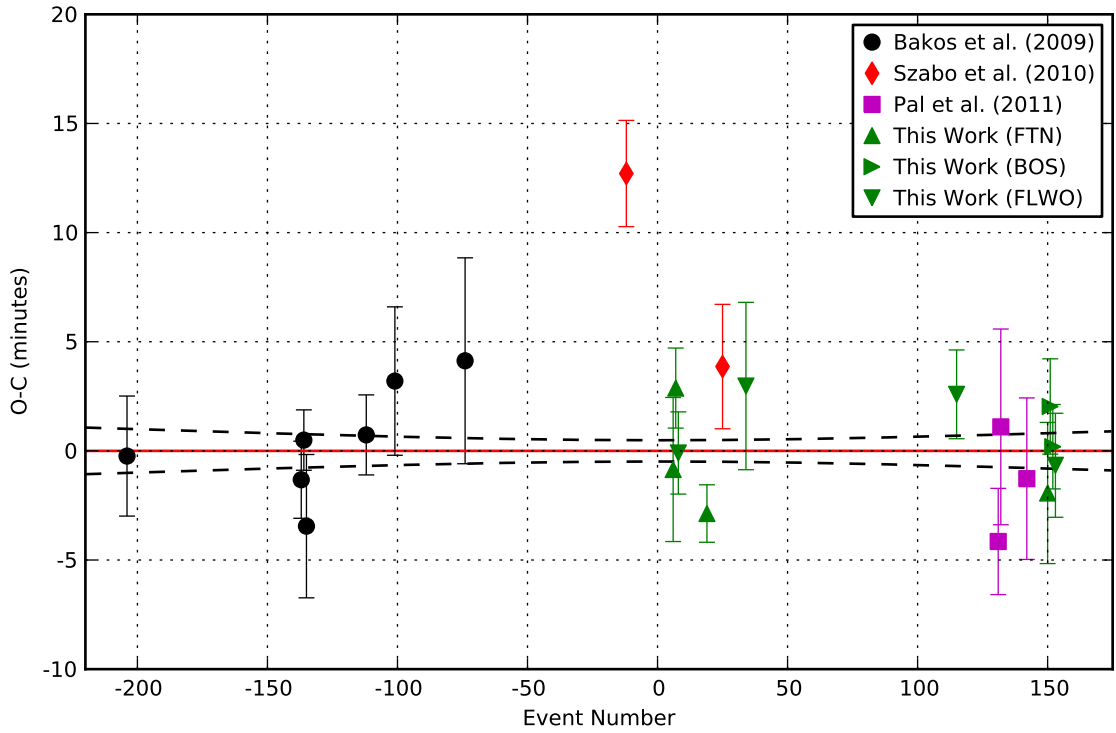}
\epsscale{0.5}
\centering
\caption{O$-$C plot including all 22 mid-transit time measurements.
The black dashed lines indicate the 1 $\sigma$ errors on the predicted mid transit times by propagating the errors on $P$ and $T_{c,0}$ from the linear fit.}
\label{oc}
\end{figure}

We present the light curve from the UT 2009 Nov 5 event in Figure \ref{szabo}, over-plotted by our model from the analysis of all light curves and shifted to the best-fitting mid transit time for the UT 2009 Nov 5 event (left), and the mid transit time from the linear ephemeris (right). A close look at the light curve residuals, presented at the bottom part of both panels, shows it includes a few features, specifically during and before ingress, and during and after egress, and these features are clearly more pronounced in the right panel. It is possible that those features have affected the estimate of the mid transit time, and their origin could be astrophysical (although HAT-P-13 is not known to be an active star), or the result of correlated noise.

We carefully examined the UT 2009 Nov 5 light curve, which consists of exposures alternating between $V$ and $R$ filters. We
measured the mid transit time for the light curve observed in each filter independently, and found that they
were both within 1.3 $\sigma$ of the mid transit time that we measured
from the combined light curve, and close to 4 $\sigma$ away from the linear ephemeris.
We then applied a completely separate analysis of the combined light curve from both filters in which the parameters that determine the shape of the light curve ($i$, $a/R_{s}$, $R_p/R_s$, and $T_{c}$) were fitted and allowed to vary freely.
The resulting fitted parameters obtained from the analysis of this single event are within 1.3 $\sigma$ from the values obtained from the Season 3 analysis.

We verified that the mid transit times we derived here for the transits observed by \cite{Bakos09} and \cite{Pal11} are consistent, within 0.7 $\sigma$, with the times derived by those authors. We measure a mid-transit time of $T_{c,n=-12}$=2455141.552706 $\pm$ 0.001700, for the UT 2009 Nov 5 transit from \citet{Szabo10} which is consistent with the value that they derive to within 0.3 $\sigma$.
Our value of $T_{c,n=25}$=2455249.447554 $\pm$ 0.001900 for the second transit obtained by \citet{Szabo10}, from UT 2010 Feb 21, differs from the published value by 1.6 $\sigma$.

As already noted, the source of the large amplitude shift in the transit timing of the UT 2009 Nov 5 event is unclear, and since our data do not include other transit events near that time it is difficult to rule out or confirm a physical process. If of astrophysical origin, it could be the result of an unusual physical process that affected also the shape of the light curve (see Figure~\ref{szabo}), and that is not seen during the other transit events analyzed here. HAT-P-13b transits observed by others close to that event, during October and November 2009 would be useful for shedding more light on this issue.

We have attempted to look for a TTV signal in our O$-$C diagram, although it does not show an excess scatter, besides the single outlier mentioned above. A parabolic fit to all 22 mid-transit times resulted in a value consistent with zero within 1 $\sigma$ for the quadratic coefficient, and with $\chi^{2}=43.62$ for 19 DoF, or $\chi_{red}^{2}=2.30$. Ignoring the UT 2009 Nov 5 outlier gives $\chi^{2}=19.31$ for 18 DoF,  or $\chi_{red}^{2}=1.07$, and the RMS of the O$-$C residuals are 210 s including the UT 2009 Nov 5 event and 144 s without that event. Therefore, we could not identify a long term trend in the transit times.

We also performed a period analysis on the residuals from the linear
ephemeris using the Lomb-Scargle (L-S) method \citep{Lomb76, Scargle82}, looking for a possible low-mass perturber \citep[e.g.,][]{Holman05}. The maximum peak of the periodogram was found to be similar
to the maximum peak of periodograms in which the data were rearranged in a random order. More quantitatively, the strongest periodogram peak was at the 52nd percentile of a sample of strongest peaks in periodograms of 10$^6$ random permutations, showing that no significant periodicity is seen in our mid transit times O$-$C residuals. The large outlier from the UT 2009 Nov 5 event of \citet{Szabo10} was not included in this L-S analysis.

Several authors have presented predicted TTV behavior of HAT-P-13b transits \citep{Bakos09, Payne11}, depending on the parameters of the second planet, HAT-P-13c. Our data put an upper limit on the maximum TTV amplitude of $\sim$150 s during
the 4 observational seasons. This reinforces the claim of \citet{Payne11} that the eccentricity of the outer planet must be less than $\sim$0.85, and the relative inclinations of the two planet's orbital planes must \emph{not} be in the range $88^{\circ}\textless i_{rel} \textless 92^{\circ}$.

Of course it could be that there is yet another, short period low mass planet lurking in the system. To that end we note that a $\sim$3 Earth mass planet orbiting at a coplanar orbit with twice the orbital period of HAT-P-13b will induce a TTV amplitude of 150 s \citep[calculated using the methods presented in][]{Pal10}, close to the detection threshold of our data. However, such a planet will induce also a 1 m s$^{-1}$ radial velocity amplitude. Therefore, the sensitivity of our transit timings to non transiting planets in a 1:2 resonance is close to that of existing radial velocity data \citep{Bakos09, Winn10}.


Our Season 2 light curves include data only from \citet{Bakos09}. Comparing our results for that season with the parameters presented by \citet{Bakos09} shows they are in good agreement, and are consistent within 0.5 $\sigma$ (see Table \ref{params}). This is an important validation of our analysis method using  the TAP software. The uncertainties we derive are larger, though, by typically 40--60\%, and 330\% for the orbital period. The latter can be explained by the additional value of the HATNet photometry in constraining the period, but the former may indicate that our uncertainties are overestimated and/or those of \citet{Bakos09} are underestimated. 
The error bars are also influenced by the fact that we did not include the UT 2008 Apr 25 event in the Season 2 analysis, since it was obtained during Season 1, although that light curve is partial and has a small impact on the parameters uncertainties.

Thanks to our large amount of data, with a much longer time span, the errors on the system light curve parameters from the analysis of all 22 light curves are smaller than those obtained by \citet{Bakos09} by 15--35\%, and more than a factor of 3 smaller for $P$.

Our separate analysis of the data from each season (see Table \ref{params}) shows a small shift, of $\approx2.5\ \sigma$ in $i$, $a/R_{s}$, and $R_p/R_s$, from Season 2 to Season 3, while the results for Seasons 3 and 4 are consistent within 1.2 $\sigma$. The transit duration 
also shows a jump of $\approx2.8\ \sigma$ between Season 2 and Season 3, but the duration for Season 2, Season 4 and all 22 light curves is consistent to within 0.5 $\sigma$. The shift in Season 3 may be influenced by the UT 2009 Nov 5 event, as the light curve (see Figure~\ref{szabo}) of this event may indicate a longer duration than other transits in our collection of data. The low significance of those shifts, and the fact that these parameters are correlated makes it difficult to draw any conclusion.
If the UT 2009 Nov 5 light curve is excluded from the Season 3 analysis, then this jump becomes slightly less significant. The values of $i$, $a/R_{s}$, and $R_{p}/R_{s}$ ($i$=81.67$\pm$0.87, $a/R_{s}$=5.29$\pm$0.31, $R_{p}/R_{s}$=0.0858$\pm$0.0023) if the UT 2009 Nov 5 event is excluded are consistent with the values of Season 3 when that event is included to within 0.4 $\sigma$, and are less than 2 $\sigma$ away from Season 2.

\section{Summary}
\label{summary}

We presented here an analysis of 22 HAT-P-13b transit light curves spanning four observational seasons, of which 10 were obtained here and 12 were previously published. Contrary to the long term TTV signal suggested by \cite{Pal11} we find that the transit times are consistent with a linear ephemeris, while we identify a single transit time, from UT 2009 Nov 5, that significantly deviates from our linear model. The nature of this single deviation is unclear. The other light curve parameters do not show a large deviation compared to those fitted to light curves of the same season. Our large data set also allows us to refine the light curve parameters and transit ephemeris, which will be useful for future observational studies of this interesting system.

This work demonstrates the use of a collaboration of ground-based 1 m class telescopes for transit timing monitoring, and that a large number of observations are required for thoroughly studying any TTV detection. This will undoubtedly be one of the goals of the future robotic 1 m class telescope networks, like LCOGT \citep[e.g.][]{Shporer10}. Unlike expensive space missions as CoRoT and Kepler, small ground-based telescopes are easily accessible and their lifetime is not limited by the durations of space missions. Therefore, they will be an important resource in studying transiting planets orbiting bright stars in the decades to come.

\acknowledgments

A.S. acknowledges support from NASA Grant Number NNX10AG02A.
M.H. and J.W. gratefully acknowledge
support from the NASA Origins program through award
NNX09AB33G.
A.P. thanks the support of the ESA grant PECS 98073, and the J\'anos Bolyai Research Scholarship of the HAS.
This paper uses observations obtained with facilities of the Las Cumbres Observatory Global Telescope.
The Byrne Observatory at Sedgwick (BOS) is operated by the Las Cumbres Observatory Global Telescope Network
and is located at the Sedgwick Reserve, a part of the University of California Natural Reserve System.
PyRAF is a product of the Space Telescope Science Institute, which is operated by AURA for NASA.

{\it Facilities:} \facility{FTN (Spectral), FLWO:1.2m (KeplerCam), LCOGT (BOS)}


\tablenum{1}
\thispagestyle{empty}
\begin{deluxetable}{ccccc}
\label{lclist}
\tablewidth{0pt} 
\tabletypesize{\small}
\tablecaption{Photometry of HAT-P-13b obtained in this work\tablenotemark{a}}
\tablehead{\colhead{BJD\_TDB} & \colhead{Relative Flux} & \colhead{Error} &  \colhead{Filter} & \colhead{Telescope\tablenotemark{b}}}

\startdata
2455193.917958 & 1.00015 & 0.00089 & $Z$ & 1 \\
2455196.836174 & 0.99512 & 0.00101 & $Z$ & 1 \\
2455199.700075 & 1.00361 & 0.00166 & $i'$ & 2 \\
2455231.842858 & 1.00464 & 0.00090 & $Z$ & 1 \\
2455275.603565 & 1.00116 & 0.00151 & $i'$  & 2 \\
2455511.818399 & 0.99610 & 0.00163 & $i'$ & 2 \\
2455613.891700 & 1.00736 & 0.00083 & $Z$ & 1 \\
2455616.787707 & 1.00060 & 0.00062 & $i'$ & 3 \\
2455619.705125 & 1.00093 & 0.00061 & $i'$ & 3 \\
2455622.690334 & 0.99150 & 0.00147 & $i'$ & 2 \\

\enddata
\tablenotetext{a}{Only a sample is given here, the full table will be available in the online version of the manuscript}
\tablenotetext{b}{Telescope code is: 1 = FTN 2.0 m, 2 = FLWO 1.2 m, 3 = BOS 0.8 m}

\end{deluxetable}
\setlength{\voffset}{0mm}

\tablenum{2} 
\thispagestyle{empty}
\begin{deluxetable}{cccccclccc}
\rotate 
\tablewidth{0pt} 
\tabletypesize{\tiny}
\tablecaption{Transit Observations of HAT-P-13b Analyzed in This Work}
\tablehead{\colhead{Date\tablenotemark{a}} & \colhead{N$_{tr}$}  & \colhead{Cycle Time} & \colhead{$\sigma_{w}$} & \colhead{$\sigma_{r}$} & \colhead{PNR\tablenotemark{b}} & \colhead{Transit Part\tablenotemark{c}} & \colhead{Filter} & \colhead{Telescope} & \colhead{Reference}\\
\colhead{(UT)} & \colhead{} & \colhead{(s)} & \colhead{(\%)} & \colhead{(\%)} & \colhead{(\% minute$^{-1}$)} & \colhead{(OIBEO)} &\colhead{} & \colhead{}}

\startdata
\multicolumn{10}{c}{Season 1 -- 2007/2008} \\
\hline
2008-04-25 & -204 & 29 & 0.19 & 0.50 & 0.14 & \ \ \ \ \ \ \ \ \,E O & $i'$ & FLWO 1.2m & \citet{Bakos09}  \\
\\ \multicolumn{10}{c}{Season 2 -- 2008/2009} \\
\hline
2008-11-06 & -137 & 29 & 0.13 & 0.41 & 0.10 & O I B & $i'$ & FLWO 1.2m & \citet{Bakos09}  \\
2008-11-09 & -136 & 29 & 0.14 & 0.56 & 0.13 & O I B E O & $i'$ & FLWO 1.2m & \citet{Bakos09}  \\
2008-11-12 & -135 & 29 & 0.18 & 0.13 & 0.18 & \ \ \ \ \,\,\,B E O & $i'$ & FLWO 1.2m & \citet{Bakos09}  \\
2009-01-18 & -112 & 29 & 0.14 & 0.14 & 0.15 & O I B E O & $i'$ & FLWO 1.2m & \citet{Bakos09}  \\
2009-02-19 & -101 & 29 & 0.18 & 0.55 & 0.14 & O I B & $i'$ & FLWO 1.2m & \citet{Bakos09}  \\
2009-05-09 & -74 & 29 & 0.15 & 0.85 & 0.14 & O I B & $i'$ & FLWO 1.2m & \citet{Bakos09}  \\
\\ \multicolumn{10}{c}{Season 3 -- 2009/2010} \\
\hline
2009-11-05 & -12 & 132 & 0.06 & 0.57 & 0.19 & O I B E O & $R\&V$ & Konkoly 1.0m & \citet{Szabo10} \\
2009-12-28 & 6 & 23 & 0.17 & 0.29 & 0.11 & O I & $Z$ & FTN 2.0m & this work \\
2009-12-31 & 7 & 22 & 0.22 & 0.58 & 0.15 & O I B & $Z$ & FTN 2.0m & this work \\
2010-01-03 & 7 & 44 & 0.13 & 0.19 & 0.14 & O I B & $i'$ & FLWO 1.2m & this work \\
2010-02-04 & 19 & 30 & 0.12 & 0.46 & 0.12 & O I B E O & $Z$ & FTN 2.0m & this work \\
2010-02-21\tablenotemark{d} & 25 & 411 & 0.09 & 0.58 & 0.43 & O I B E O & $R$ & Konkoly 0.6m & \citet{Szabo10} \\
2010-03-20 & 34 & 34 & 0.15 & 0.38 & 0.12 & O I B & $i'$ & FLWO 1.2m & this work \\
\\ \multicolumn{10}{c}{Season 4 -- 2010/2011} \\
\hline
2010-11-11 & 115 & 39 & 0.16 & 0.53 & 0.14 & O I B E O & $i'$ & FLWO 1.2m & this work \\
2010-12-27 & 131 & 39 & 0.19 & 0.75 & 0.17 & O I B E O & $I$ & Konkoly 0.6m & \citet{Pal11} \\
2010-12-30 & 132 & 63 & 0.08 & 1.29 & 0.17 & O I B & $R$ & Konkoly 1.0m & \citet{Pal11} \\
2011-01-28 & 142 & 28 & 0.17 & 1.44 & 0.15 & O I B E & $R$ & Konkoly 1.0m & \citet{Pal11} \\
2011-02-21 & 150 & 31 & 0.27 & 0.55 & 0.23 & O I B E & $Z$ & FTN 2.0m & this work \\
2011-02-24 & 151 & 62 & 0.10 & 0.19 & 0.11 & O I B & $i'$ & BOS 0.8m & this work \\
2011-02-27 & 152 & 94 & 0.11 & 0.37 & 0.15 & O I B E O & $i'$ & BOS 0.8m & this work \\
2011-03-02 & 153 & 34 & 0.11 & 0.65 & 0.10 &  \ \ \ \ \,\,\,B E O & $i'$ & FLWO 1.2m & this work \\

\\
\enddata
\tablenotetext{a}{UT date at start of observation.}
\tablenotetext{b}{Photometric noise rate, calculated as RMS/$\sqrt{\Gamma}$, where RMS is the scatter in the light curve residuals and $\Gamma$ is the median number of cycles (exposure time and dead time) per minute.}
\tablenotetext{c}{OIBEO for Out-of-transit before ingress, Ingress, flat Bottom, Egress, and Out-of-transit after egress respectively.}
\tablenotetext{d}{\rm The long cycle time (and resulting PNR) is the result of ignoring the $V$ band data of this event. Observations were originally
taken while alternating between $V$ and $R$ filters, but the V band data were plagued by large systematics and ignored in the analysis of \citet{Szabo10}.}
\label{obstable}
\end{deluxetable}
\setlength{\voffset}{0mm}

\clearpage


\begin{deluxetable}{cccccc}
\rotate
\tablewidth{0pt}
\tablenum{3}
\tabletypesize{\small}
\tablecaption{Light Curve Parameters}
\tablehead{
\colhead {} & \colhead{$T_{c,0}$} & \colhead{$P$} & \colhead{$i$} & \colhead{$a/R_{s}$} & \colhead{$R_p/R_{s}$ }\\
& \colhead{(BJD\_TDB)} & \colhead{(days)} & \colhead{(deg)}
}
\startdata
{\bf All} & {\bf2455176.53880 $\pm$ 0.00034} & {\bf2.9162430 $\pm$ 0.0000030} & {\bf82.45 $\pm$ 0.46} &  {\bf5.52 $\pm$ 0.17} & {\bf0.0855 $\pm$ 0.0011} \\
Season 2 -- 2008/2009 & 2454779.92895 $\pm$ 0.00072 &  2.916305 $\pm$ 0.000033 &  83.4 $^{+1.0}_{-0.88}$ & 5.86 $^{+0.41}_{-0.34}$ & 0.0838 $\pm$ 0.0019\\
Season 3 -- 2009/2010 &  2455231.9464 $\pm$ 0.0012 & 2.915952 $\pm$ 0.000063 & 81.16 $\pm$ 0.70 & 5.00 $\pm$ 0.23 & 0.0882 $\pm$ 0.0020 \\
Season 4 -- 2010/2011 & 2455619.80708 $\pm$ 0.00085 & 2.916203 $\pm$ 0.000048 & 82.05 $\pm$ 0.93  & 5.37 $\pm$ 0.36 & 0.0857 $\pm$ 0.0021 \\
\citet{Bakos09} & 2454779.92979 $ \pm$ 0.00038 & 2.916260 $\pm$ 0.000010 & 83.4 $\pm$ 0.6 & 5.84 $\pm$ 0.26 & 0.0844 $\pm$ 0.0013 \\
\enddata
\label{params}
\end{deluxetable}


\tablenum{4}
\thispagestyle{empty}
\begin{deluxetable}{ccccc}
\tablewidth{0pt} 
\tabletypesize{\small}
\tablecaption{Mid-transit times of HAT-P-13b for the 22 light curves analyzed here}
\tablehead{\colhead{N$_{tr}$} & \colhead{T$_{c}$} & \colhead{$\sigma_{T_{c}}$} &  \colhead{O$-$C} & \colhead{O$-$C/$\sigma_{T_{c}}$}\\
\colhead{} & \colhead{(BJD\_TDB)} & \colhead{(s)} & \colhead{(s)}}

\startdata
-204 & 2454581.625065 & 165 & -13 & -0.08\\
-137 & 2454777.012591 & 106 & -79 & -0.74\\
-136 & 2454779.930096 & 83 & 30 & 0.36\\
-135 & 2454782.843600 & 197 & -206 & -1.05\\
-112 & 2454849.920092 & 110 & 44 & 0.40\\
-101 & 2454882.000478 & 204 & 192 & 0.94\\
-74 & 2454960.739683 & 283 & 248 & 0.88\\

-12 & 2455141.552706 & 146 & 762 & 5.22\\
6 & 2455194.035662 & 198 & -51 & 0.26\\
7 & 2455196.954496 & 110 & 173 & 1.57\\
8 & 2455199.868674 & 113 & -6 & -0.05\\
19 & 2455231.945420 & 79 & -172 & -2.18\\
25 & 2455249.447554 & 171 & 232 & 1.36\\
34 & 2455275.693121 & 230 & 178 & 0.78\\

115 & 2455511.908538 & 122 & 155 & 1.27\\
131 & 2455558.563744 & 146 & -249 & -1.71\\
132 & 2455561.483633 & 269 & 66 & 0.25\\
142 & 2455590.644415 & 222 & -76 & -0.34\\
150 & 2455613.973903 & 194 & -116 & -0.60\\
151 & 2455616.892899 & 131 & 122 & 0.93\\
152 & 2455619.807862 & 116 & 11 & 0.10\\
153 & 2455622.723514 & 143 & -39 & -0.28\\
\enddata

\label{midtable}
\end{deluxetable}
\setlength{\voffset}{0mm}


\bibliographystyle{apj}
\bibliography{references}
\end{document}